\documentclass[preprint,showpacs,showkeys,preprintnumbers,amsmath,amssymb]{revtex4}
 \usepackage{dcolumn}% Align table columns on decimal point
 \usepackage{bm}% bold math

\begin{document}

 \title{ Entropy of Kaluza-Klein Black Hole
  from Kerr/CFT Correspondence }

 \author{ Ran Li }

 \thanks{Electronic mail: liran05@lzu.cn}

 \affiliation{ Center of Theoretical Nuclear Physics,
 National Laboratory of Heavy Ion Accelerator, Lanzhou 730000, Gansu, China}
 \affiliation{Institute of Modern Physics, Chinese Academy of Sciences,
  Lanzhou, 730000, Gansu, China}

 \author{ Ming-Fan Li }

 \thanks{Electronic mail: limf07@lzu.cn}

 \author{ Ji-Rong Ren }

 \thanks{Electronic mail: renjr@lzu.edu.cn}

 \affiliation{Institute of Theoretical Physics, Lanzhou University, Lanzhou, 730000, Gansu, China}

 \begin{abstract}

 We extend the recently proposed Kerr/CFT correspondence
 to examine the dual conformal field theory
 of four dimensional Kaluza-Klein black hole
 in Einstein-Maxwell-Dilaton theory.
 For the extremal Kaluza-Klein black hole,
 the central charge and temperature of the dual conformal field
 are calculated following the approach of
 Guica, Hartman, Song and Strominger. Meanwhile,
 we show that the microscopic entropy
 given by the Cardy formula agrees with
 Bekenstein-Hawking entropy of extremal Kaluza-Klein
 black hole.
 For the non-extremal case, by studying the near-region wave
 equation of a neutral massless scalar field,
 we investigate the hidden conformal symmetry of Kaluza-Klein
 black hole, and find the
 left and right temperatures of the dual
 conformal field theory.
 Furthermore, we find that the entropy of non-extremal
 Kaluza-Klein black hole is reproduced
 by Cardy formula.

 \end{abstract}

 \pacs{}

 \keywords{Kerr/CFT correspondence, Kaluza-Klein black hole,
  Bekenstein-Hawking entropy}

 \maketitle

 \section{introduction}

 The recently proposed Kerr/CFT correspondence \cite{GTSS}
 states that quantum gravity in the region
 very near the horizon of an extreme Kerr black hole
 with proper boundary conditions
 is holographically dual to a two-dimensional
 chiral conformal field theory with the central
 charge proportional to angular momentum.
 It is shown that the macroscopic Bekenstein-Hawking
 entropy of extremal Kerr black hole can be reproduced
 by the microscopic entropy of dual conformal field theory
 via Cardy formula.
 The method employed by Guica, Hartman, Song and Strominger(GHSS)
 in \cite{GTSS} is very similar to the approach of
 Brown and Henneaux in \cite{Brown86},
 where the $AdS_3$ background
 is replaced by the near-horizon extremal Kerr(NHEK) geometry
 previously obtained in \cite{BH}.
 This method has
 been generalized to calculate the entropies of extremal black holes in a
 lot of theories such as the Einstein theory, string theory, and
 supergravity theory, as well as those solutions in diverse
 dimensions\cite{HMNS,lu,Azeyanagi,Nakayama,chow,isono,azeyanagi1,
 peng,loran,chen,ghezelbash,lumei,compere,hotta,astefanesei,
 garousi,ghezelbash1,azeyanaqi2,krishnan1,
 wutian,wen,cao,matsuo,barnes,annoinos,rasmussen,peng1,rasmussen2,
 mei,Castrolarsen,chensz,Soltanpanahi,Bredberg,Hartman,Cvetic}.

 Very recently, Castro, Maloney and Strominger (CMS)
 in a remarkable paper \cite{Castro} show that there exists a
 hidden conformal symmetry for four dimensional non-extremal Kerr Black hole
 by studying the near-region wave equation of a
 massless scalar field.
 Interestingly, the hidden conformal symmetry is not derived from
 the conformal symmetry of spacetime geometry itself.
 It is also shown that
 microscopic entropy
 computed by Cardy formula agrees exactly with
 the macroscopic Berenstein-Hawking
 entropy of non-extremal Kerr black hole.
 These observations suggest that non-extremal
 Kerr black hole is holographically dual to a two-dimensional
 conformal field theory with non-zero left and right
 temperatures.
 Subsequently, with the idea of hidden conformal
 symmetry of non-extremal black holes,
 Krishnan \cite{Krishnan} extended this method to investigate five-dimensional
 black holes in string theory. Furthermore, the non-extremal
 uplifted 5D Reissner-Nordstrom black hole was investigated by
 Chen and Sun \cite{Chensun}, and the hidden conformal
 symmetry of Kerr-Newman black hole was
 studied by Wang et al. in \cite{Wang} and Chen et al. in \cite{chenlong}.

 In this paper, We extend the Kerr/CFT correspondence
 to examine the dual conformal field theory
 of four dimensional Kaluza-Klein black hole
 in Einstein-Maxwell-Dilaton theory.
 For the extremal Kaluza-Klein black hole,
 we firstly perform a coordinates transformation
 to find the near horizon extremal geometry.
 Then, by employing the approach of GHSS,
 the central charge and the left and right temperatures
 of dual conformal field are calculated.
 Finally, we find the microscopic entropy
 calculated by using Cardy formula agrees with
 Bekenstein-Hawking entropy of the extremal Kaluza-Klein
 black hole.
 For the non-extremal case,
 we investigate the hidden conformal symmetry of Kaluza-Klein
 black hole by studying the near-region wave
 equation of a neutral massless scalar
 field in this background, and find the
 left and right temperatures of dual
 conformal field theory.
 Furthermore, the entropy of non-extremal
 Kaluza-Klein black hole is reproduced
 by using Cardy formula.

 It should be noted that the holographic dual of
 the extremal uplifted 5D Kaluza-Klein black hole
 has been investigated by Azeyanagi et.al. in \cite{Azeyanagi}.
 They show that the central charge of
 dual conformal field is $c=12J$, where $J$ is the angular
 momentum of 5D Kaluza-Klein black hole.
 In this paper, we treat the Kaluza-Klein
 black hole as a exact solution
 of Einstein-Maxwell-Dilaton theory.
 It seems that there exists a non-vanishing
 contributions to the central charge from
 gauge field and Dilaton field.
 Fortunately, an explicit calculation
 given by Compere et al. in \cite{compere} shows that
 the central charge receives no contribution from the
 non-gravitational fields, i.e.
 only the Einstein-Hilbert Lagrangian contributes
 to the value of the central charge.
 So, it is sufficient to only calculate the gravitational
 field contribution to the central charge
 for four dimensional Kaluza-Klein black hole
 in the present situation.
 Our result presented in Sec.III
 agrees with the observation in \cite{Azeyanagi},
 which confirms this viewpoint.

 This paper is organized as follows.
 In section II, we give a brief review of four dimensional Kaluza-Klein black
 hole. In section III,
 we calculate the central charge and the left and right
 temperatures of the dual conformal field theory
 for the extremal Kaluza-Klein black hole,
 and find the microscopic entropy of the dual CFT.
 In section IV, we study the hidden conformal
 symmetry of the non-extremal Kaluza-Klein
 black hole by analysing the near-region wave
 equation of a neutral massless scalar
 field. Furthermore,
 the microscopic entropy of dual
 CFT with non-zero left and right
 temperatures are obtained.
 The last section is devoted to discussion.

 \section{Kaluza-Klein black hole}

 In this section, we will give a brief review of four dimensional Kaluza-Klein black
 hole. Kaluza-Klein black hole solution is derived by a dimensional reduction
 of the boosted five-dimensional Kerr solution to four dimensions.
 It is also an exact solution of Einstein-Maxwell-Dilaton action.
 The metric is explicitly given by\cite{KK,KK1}
 \begin{eqnarray}
 ds^2&=&-\frac{1-Z}{B}dt^2-\frac{2aZ\textrm{sin}^2\theta}{B\sqrt{1-\nu^2}}dtd\varphi+\frac{B\Sigma}{\Delta}dr^2\nonumber\\
 &&+\left[B(r^2+a^2)+a^2\textrm{sin}^2\theta\frac{Z}{B}\right]\sin^2\theta d\varphi^2+B\Sigma d\theta^2\;,
 \end{eqnarray}
 where
 \begin{eqnarray}
 Z&=&\frac{2\mu r}{\Sigma}\;,\nonumber\\
 B&=&\sqrt{1+\frac{\nu^2 Z}{1-\nu^2}}\;,\nonumber\\
 \Sigma&=&r^2+a^2\textrm{cos}^2\theta\;,\nonumber\\
 \Delta&=&r^2-2\mu r+a^2\;.
 \end{eqnarray}
 The dilaton field and gauge potential are given by respectively
 \begin{eqnarray}
 \phi&=&-\frac{\sqrt{3}}{2}\textrm{ln}B\;,\nonumber\\
 A&=&\frac{\nu }{2(1-\nu^2)}\frac{Z}{B^2}dt-\frac{a\nu
 \textrm{sin}^2\theta}{2\sqrt{1-\nu^2}}\frac{Z}{B^2}d\varphi\;.
 \end{eqnarray}
 The physical mass $M$, the charge $Q$, and the angular momentum
 $J$ are expressed with the boost parameter $\nu$, mass parameter
 $\mu$, and specific angular momentum $a$ as
 \begin{eqnarray}
 &&M=\mu\Big[1+\frac{\nu^2}{2(1-\nu^2)}\Big]\;,\nonumber\\
 &&Q=\frac{\mu\nu}{1-\nu^2}\;,\nonumber\\
 &&J=\frac{\mu a}{\sqrt{1-\nu^2}}\;.
 \end{eqnarray}
 The outer and inner horizons are respectively given by
 \begin{equation}
 r_\pm=\mu\pm\sqrt{\mu^2-a^2}\;.
 \end{equation}
 Hawking temperature, entropy and angular velocity of the event horizon
 are respectively given by
 \begin{eqnarray}
 T_H&=&\frac{\sqrt{1-\nu^2}}{2\pi}\frac{\mu^2-a^2}{r_+^2+a^2}\;,\nonumber\\
 \Omega_H&=&\frac{a\sqrt{1-\nu^2}}{r_+^2+a^2}\;,\nonumber\\
 S&=&2\pi\frac{\mu}{\sqrt{1-\nu^2}}(\mu+\sqrt{\mu^2-a^2})\;.
 \end{eqnarray}
 The extremality condition is $\mu=a$,
 and the entropy at extremality is
 \begin{eqnarray}
 S(T_H=0)=\frac{2\pi\mu^2}{\sqrt{1-\nu^2}}\;.
 \end{eqnarray}
 In the following two sections,
 we will try to reproduce the
 Bekenstein-Hawking entropies of the extremal and non-extremal
 Kaluza-Klein black hole via Cardy formula
 of the dual conformal field.

 \section{Entropy of Extremal Kaluza-Klein black hole from Kerr/CFT dual }

 In this section, our purpose is to derive Bekenstein-Hawking entropy
 of extremal Kaluza-Klein black hole via the extremal
 Kerr/CFT correspondence.

 We now try to explore the near-horizon geometry
 of extremal Kaluza-Klein black hole. To do
 so, we need to perform the following coordinate transformations
 \begin{eqnarray}
 r&=&a+\epsilon r_0 \hat{r}\;,\nonumber\\
 t&=&\frac{r_0 \hat{t}}{\epsilon}\;,\nonumber\\
 \varphi&=&\hat{\varphi}+\frac{\sqrt{1-\nu^2}}{2a}\frac{r_0
 \hat{t}}{\epsilon}\;,
 \end{eqnarray}
 with the parameter $r_0^2=\frac{2a^2}{\sqrt{1-\nu^2}}$.
 After taking the $\epsilon\rightarrow 0$ limit,
 the near-horizon geometry of an extremal Kaluza-Klein black
 hole reads
 \begin{eqnarray}
 ds^2&=&\hat{B}a^2(1+\cos^2\theta)\left(-\hat{r}^2d\hat{t}^2+
  \frac{d\hat{r}^2}{\hat{r}^2}+d\theta^2\right)\nonumber\\
  &&+\frac{4a^2}{\hat{B}(1-\nu^2)}\frac{\sin^2\theta}{1+\cos^2\theta}
  \left(d\hat{\varphi}+\hat{r}d\hat{t}\right)^2\;,
 \end{eqnarray}
 with
 \begin{eqnarray}
 \hat{B}=\left[1+\frac{2\nu^2}{(1-\nu^2)(1+\cos^2\theta)}\right]^{\frac{1}{2}}\;.
 \end{eqnarray}

 We now employ the Brown-Henneaux approach to
 find the central charge of the dual holographic
 conformal field theory description
 of an extremal Kaluza-Klein black hole.
 As explained in the introduction,
 the calculation carried by Compere et al in \cite{compere}
 indicates that, in Einstein-Maxwell-Dilaton theory
 with topological terms in four and five dimensions,
 the central charge receives no contribution from the
 non-gravitational fields. To find the
 central charge of the dual conformal field for
 four dimensional Kaluza-Klein black hole
 in Einstein-Maxwell-Dilaton theory, for simplicity,
 it is sufficient to only calculate the gravitational
 field contribution.

 It is important to impose the appropriate boundary conditions at spatial
 infinity and find the asymptotical symmetry group that preserves
 these boundary conditions. We choose the boundary conditions
 \begin{eqnarray}
 \left(
   \begin{array}{cccc}
     h_{\hat{t}\hat{t}}=\mathcal{O}(\hat{r}^2)
     & h_{\hat{t}\hat{\varphi}}=\mathcal{O}(1)
     & h_{\hat{t}\theta}=\mathcal{O}(\frac{1}{\hat{r}})
     & h_{\hat{t}\hat{r}}=\mathcal{O}(\frac{1}{\hat{r}^2}) \\
      & h_{\hat{\varphi}\hat{\varphi}}=\mathcal{O}(1)
      & h_{\hat{\varphi}\theta}=\mathcal{O}(\frac{1}{\hat{r}})
      & h_{\hat{\varphi}\hat{r}}=\mathcal{O}(\frac{1}{\hat{r}}) \\
      &  & h_{\theta\theta}=\mathcal{O}(\frac{1}{\hat{r}})
      & h_{\theta\hat{r}}=\mathcal{O}(\frac{1}{\hat{r}^2})  \\
      &  &  & h_{\hat{r}\hat{r}}=\mathcal{O}(\frac{1}{\hat{r}^3}) \\
   \end{array}
 \right)
 \end{eqnarray}
 where $h_{\mu\nu}$ is the metric deviation
 from the near horizon geometry.
 The diffeomorphism symmetry that
 preserves such a boundary condition is generated by the
 vector field
 \begin{eqnarray}
 \zeta=\epsilon(\hat{\varphi})\frac{\partial}{\partial\hat{\varphi}}
 -\hat{r}\epsilon'(\hat{\varphi})\frac{\partial}{\partial\hat{r}}\;,
 \end{eqnarray}
 where $\epsilon(\hat{\varphi})$ is an arbitrary
 smooth periodic function of the coordinate $\hat{\varphi}$.
 It is convenient to define $\epsilon_n(\hat{\varphi})=-e^{-in\hat{\varphi}}$
 and $\zeta_n=\zeta(\epsilon_n)$, where $n$ are integers.
 Then the asymptotic symmetry group is generated by
 \begin{eqnarray}
 \zeta_n=-e^{-in\hat{\varphi}}\frac{\partial}{\partial\hat{\varphi}}
 -in\hat{r}e^{-in\hat{\varphi}}\frac{\partial}{\partial\hat{r}}\;,
 \end{eqnarray}
 which obey the Virasoro algebra
 \begin{eqnarray}
 i[\zeta_m,\zeta_n]=(m-n)\zeta_{m+n}\;.
 \end{eqnarray}
 Each diffeomorphism $\zeta_n$ is associated
 to a conserved charge defined by\cite{barnich}
 \begin{eqnarray}
 Q_{\zeta}=\frac{1}{8\pi}\int_{\partial\Sigma} k_{\zeta}\;,
 \end{eqnarray}
 where $\partial\Sigma$ is a spatial slice, and
 2-form $k_{\zeta}$ is defined as
 \begin{eqnarray}
 k_{\zeta}[h,g]&=&\frac{1}{2}\left[
 \zeta_\nu\nabla_\mu h-\zeta_\nu\nabla_\sigma h_\mu^{\;\;\sigma}
 +\zeta_\sigma\nabla_\nu h_\mu^{\;\;\sigma}
 +\frac{1}{2}h\nabla_\nu\zeta_\mu\right.\nonumber\\
 &&\left.-h_\nu^{\;\;\sigma}\nabla_\sigma\zeta_\mu
 +\frac{1}{2}h_{\nu\sigma}\left(\nabla_\mu\zeta^\sigma
 +\nabla^\sigma\zeta_\mu\right)
 \right]*\left(dx^\mu\wedge dx^\nu\right)\;,
 \end{eqnarray}
 where $*$ denotes the Hodge dual.
 The Dirac brackets of the conserved charges
 are just the common forms of the Virasoro algebras
 with central terms
 \begin{eqnarray}
 \frac{1}{8\pi}\int_{\partial\Sigma} k_{\zeta_m}[\mathcal{L}_{\zeta_n}g,g]
 =-\frac{i}{12}c(m^3+\alpha m)\delta_{m+n,0}\;,
 \end{eqnarray}
 where $c$ denote the central charge
 corresponding to the diffeomorphism
 and $\alpha$ is a trial constant.
 Evaluating the integral for the case of
 the near-horizon extremal Kaluza-Klein metric,
 we find the central charge
 \begin{eqnarray}
 c=\frac{12\mu a}{\sqrt{1-\nu^2}}\;.
 \end{eqnarray}

 This result exactly agrees with
 the one obtained by Azeyanagi et al. in \cite{Azeyanagi}.
 It should be noted that
 the central charge can also be written
 as $c=12J$. This relation between
 the central charge and angular momentum
 is just the same as that for Kerr
 black hole\cite{GTSS} and other examples
 of the Extremal Kerr/CFT dual.

 After obtaining the central charge of the extremal Kaluza-Klein
 black hole, we now begin to get its CFT entropy. To get this, we have to
 calculate the generalized temperature with respect to the
 Frolov-Thorne vacuum.
 We consider the quantum field with
 eigenmodes of the asymptotic energy $\omega$ and angular momentum
 $m$, which are given by the following form
 \begin{eqnarray}
 e^{-i\omega t+im\varphi}=
 e^{-i\left(\omega-\frac{m\sqrt{1-\nu^2}}{2a}\right)
 \frac{r_0}{\epsilon}\hat{t}+im\hat{\varphi}}
 =e^{-in_R\hat{t}+in_L\hat{\varphi}}\;,
 \end{eqnarray}
 with
 \begin{eqnarray}
 n_R=\left(\omega-\frac{m\sqrt{1-\nu^2}}{2a}\right)
 \frac{r_0}{\epsilon}\;,\;\;\;n_L=m\;.
 \end{eqnarray}
 The correspondence Boltzmann factor is of the form
 \begin{eqnarray}
 e^{-\frac{\omega-m\Omega}{T_H}}=e^{-\frac{n_R}{T_R}-\frac{n_L}{T_L}}\;,
 \end{eqnarray}
 where the left and right temperatures are given by
 \begin{eqnarray}
 T_R&=&\frac{r_0}{\epsilon}T_H\;,\nonumber\\
 T_L&=&\frac{T_H}{\frac{\sqrt{1-\nu^2}}{2a}-\Omega_H}\;.
 \end{eqnarray}
 In the extremal limit $\mu\rightarrow a$,
 the left and right temperatures reduce to
 \begin{eqnarray}
 T_R=0\;,\;\;\;T_L=\frac{1}{2\pi}\;.
 \end{eqnarray}
 According to the Cardy formula the entropy for a unitary CFT,
 we can obtain the microscopic entropy of the extremal Kaluza-Klein
 black hole
 \begin{eqnarray}
 S_{CFT}=\frac{\pi^2}{3}cT_L=\frac{2\pi\mu a}{1-\nu^2}\;,
 \end{eqnarray}
 which precisely agrees with the Bekenstein-Hawking entropy.

 \section{entropy of non-extremal Kaluza-Klein
  black hole from hidden conformal symmetry}

 In this section, we extend the analysis of hidden
 conformal symmetry for the Kerr black hole to
 the Kaluza-Klein balck hole.

 Let us consider the Klein-Gordon equation for
 the neutral massless scalar field in the background
 of the Kaluza-Klein black hole
 \begin{eqnarray}
 \frac{1}{\sqrt{-g}}\partial_\mu\left
 (\sqrt{-g}g^{\mu\nu}\partial_\nu\right)\Phi=0\;.
 \end{eqnarray}
 The scalar field wave function can be expanded in
 eigenmodes as
 \begin{eqnarray}
 \Phi=e^{-i\omega t+im\varphi}\Phi(r,\theta)\;,
 \end{eqnarray}
 where $\omega$ and $m$ are the quantum numbers.
 Then the scalar field wave equation can be separated into
 the spheroidal equation and the radial equation
 \begin{eqnarray}
 \left[\frac{1}{\sin\theta}\frac{d}{d\theta}
 \left(\sin\theta\frac{d}{d\theta}\right)
 -\left(\frac{m^2}{\sin^2\theta}+a^2\omega^2\sin^2\theta\right)\right]
 S(\theta)&=&-\lambda S(\theta)\;,\nonumber\\
 \left[\partial_r\Delta\partial_r+
 \frac{R^4(r)(\omega-m\Omega(r))^2}{\Delta}
 +\frac{1}{R^4(r)}m^2a^2\left(r^2+a^2+\frac{2\mu r}{1-\nu^2}\right)
 \right]\Psi&=&\lambda\Psi\;,
 \end{eqnarray}
 where $\lambda$ is the separation
 constant and
 \begin{eqnarray}
 \Omega(r)&=&\frac{2\mu
 r}{\sqrt{1-\nu^2}}\frac{a}{R^4(r)}\;,\nonumber\\
 R^4(r)&=&(r^2+a^2)\left(r^2+a^2+\frac{\nu^2}{1-\nu^2}2\mu
 r\right)\;.
 \end{eqnarray}
 The radial equation can also be rewritten as
 \begin{eqnarray}
 \left[\partial_r\Delta\partial_r
 +\frac{\left(2\mu r_+\omega/\sqrt{1-\nu^2}-am\right)^2}{(r-r_+)(r_+-r_-)}
 -\frac{\left(2\mu r_-\omega/\sqrt{1-\nu^2}-am\right)^2}{(r-r_-)(r_+-r_-)}
 \right.\nonumber\\
 +\left.\left(r^2+a^2+\frac{2\mu(2\mu-r)}{1-\nu^2}\right)\omega^2
 \right]\Psi=\lambda\Psi\;.
 \end{eqnarray}
 Following the argument of CMS,
 we also consider the same near-region,
 which is defined by the condition
 \begin{eqnarray}
 \omega\mu\ll 1\;,\;\;\;\omega r\ll 1\;.
 \end{eqnarray}
 In the near-region, the angular equation
 reduces to the standard Laplacian on
 the 2-sphere with the separation
 constants taking values
 \begin{eqnarray}
 \lambda=l(l+1)\;.
 \end{eqnarray}
 And the radial equation can be simplified as
 \begin{eqnarray}
 \left[\partial_r\Delta\partial_r
 +\frac{\left(2\mu r_+\omega/\sqrt{1-\nu^2}-am\right)^2}{(r-r_+)(r_+-r_-)}
 -\frac{\left(2\mu r_-\omega/\sqrt{1-\nu^2}-am\right)^2}{(r-r_-)(r_+-r_-)}
  \right]\Psi=l(l+1)\Psi\;.
 \end{eqnarray}
 The above equation can be solved by hypergeometric functions.
 As hypergeometric functions
 transform in representations of SL(2,R), this suggests the
 existence of a hidden conformal symmetry.
 Now we will show that the radial equation
 can also be obtained by using of the SL(2,R) Casimir operator.

 Introducing the coordinates
 \begin{eqnarray}
 w^+&=&\sqrt{\frac{r-r_+}{r-r_-}}e^{2\pi T_R\varphi}\;,\nonumber\\
 w^-&=&\sqrt{\frac{r-r_+}{r-r_-}}e^{2\pi
 T_L\varphi-2n_L t}\;,\\
 y&=&\sqrt{\frac{r_+-r_-}{r-r_-}}e^{\pi(T_L+T_R)\varphi-n_L
 t}\;,\nonumber
 \end{eqnarray}
 with
 \begin{eqnarray}
 T_R=\frac{r_+-r_-}{4\pi a}\;,\;\;
 T_L=\frac{r_++r_-}{4\pi a}\;,\;\;
 n_L=\frac{\sqrt{1-\nu^2}}{4\mu}\;.
 \end{eqnarray}
 Then we can locally define the vector fields
 \begin{eqnarray}
 H_1&=&i\partial_+\;,\nonumber\\
 H_0&=&i(w^+\partial_++\frac{1}{2}y\partial_y)\;,\\
 H_{-1}&=&i(w^{+2}\partial_++w^+y\partial_y-y^2\partial_-)\;,\nonumber
 \end{eqnarray}
 and
 \begin{eqnarray}
 \bar{H}_1&=&i\partial_-\;,\nonumber\\
 \bar{H}_0&=&i(w^-\partial_-+\frac{1}{2}y\partial_y)\;,\\
 \bar{H}_{-1}&=&i(w^{-2}\partial_-+w^-y\partial_y-y^2\partial_+)\;,\nonumber
 \end{eqnarray}
 These vector fields obey the SL(2, R) Lie algebra
 \begin{eqnarray}
 [H_0,H_{\pm 1}]=\mp iH_{\pm 1}\;,\;\;[H_{-1},H_1]=-2iH_0\;,
 \end{eqnarray}
 and similarly for $(\bar{H}_0,\bar{H}_{\pm1})$.
 The SL(2, R) quadratic Casimir operator is
 \begin{eqnarray}
 \mathcal{H}^2=\mathcal{\bar{H}}^2&=&-H_0^2+\frac{1}{2}(H_1H_{-1}+H_{-1}H_1)\nonumber\\
 &=&\frac{1}{4}(y^2\partial_y^2-y\partial_y)+y^2\partial_+\partial_-\;.
 \end{eqnarray}
 In terms of the $(t,r,\varphi)$ coordinates,
 the SL(2, R) quadratic Casimir operator becomes
 \begin{eqnarray}
 \mathcal{H}^2=\partial_r\Delta\partial_r
 -\frac{\left(2\mu r_+\partial_t/\sqrt{1-\nu^2}+a\partial_\varphi\right)^2}{(r-r_+)(r_+-r_-)}
 +\frac{\left(2\mu
 r_-\partial_t/\sqrt{1-\nu^2}+a\partial_\varphi\right)^2}{(r-r_-)(r_+-r_-)}\;.
  \end{eqnarray}
 So the near region wave equation can be written as
 \begin{eqnarray}
 \mathcal{H}^2\Phi=\mathcal{\bar{H}}^2\Phi=l(l+1)\Phi\;,
 \end{eqnarray}
 and the conformal weights of dual operator of the massless field $\Phi$ should be
 \begin{eqnarray}
 (h_L,h_R)=(l,l)\;.
 \end{eqnarray}

 Now, we want to calculate the
 microscopic entropy of the dual CFT,
 and compare it with the Bekenstein-Hawking entropy
 of the non-extremal Kaluza-Klein black hole.
 For the extremal case, the central charges can be
 derived from an analysis of the asymptotic
 symmetry group as we did in the last section.
 However, we did not know how to extend this
 calculation away from extremality. As did in \cite{Castro},
 we will simply assume that the
 conformal symmetry found here connects smoothly to
 that of the extreme limit and
 the central charge still keeps the same as the extremal case,
 which is given by  Eq.(18).
 The microscopic entropy of the dual CFT
 can be computed by the Cardy formula
 \begin{eqnarray}
 S_{CFT}=\frac{\pi^2}{3}(c_LT_L+c_RT_R)
 =\frac{2\pi\mu}{\sqrt{1-\nu^2}}(\mu+\sqrt{\mu^2-a^2})\;,
 \end{eqnarray}
 which matches with the black hole Bekenstein-Hawking entropy.

 \section{conclusion}

 In this paper, we have extend the recently proposed Kerr/CFT correspondence
 to examine the dual conformal field theory
 of the Kaluza-Klein black hole. Firstly,
 for the extremal Kaluza-Klein black hole,
 we have calculated
 the central charge and temperature of the dual conformal field
 by employing the approach of GHSS.
 It is shown that the microscopic entropy
 calculated by using Cardy formula agrees with the
 Bekenstein-Hawking entropy of the extremal Kaluza-Klein
 black hole. Then, for the non-extremal case,
 we have investigated the hidden conformal symmetry of Kaluza-Klein
 black hole by studying the near-region wave
 equation of a neutral massless scalar
 field, and found the
 left and right temperatures of the dual
 conformal field theory.
 Furthermore, the entropy of non-extremal
 Kaluza-Klein black hole is reproduced
 by using Cardy formula.

 The results of this paper totally
 support the arguments made in Ref.\cite{Castro},
 which suggests that, for the general rotating black hole,
 even away from extremality,
 there is a dual two dimensional conformal field theory
 with the left and right excitated modes.
 Up to now, only the hidden conformal symmetry of the near-region
 scalar field wave equation is studied.
 It would also be interesting to investigate whether
 the hidden conformal symmetry can be
 obtained by studying the near-region wave
 equation of high spin field.

 \section*{Acknowledgement}

 This work was supported by the Cuiying Programme of Lanzhou
 University (225000-582404) and the Fundamental Research Fund for
 Physics and Mathematic of Lanzhou University(LZULL200911).

 \end{document}